\documentclass[onecolumn,11pt]{elsarticle}

\usepackage{epstopdf}
\usepackage{epsfig}
\usepackage{amssymb}
\usepackage{color}
\usepackage{bm}
\usepackage{amsfonts}
\usepackage{lineno,hyperref}
\usepackage{array}
\usepackage[a4paper, total={7.0in, 9.5in}]{geometry}

\begin{document}

\begin{frontmatter}

\title{Lorentzian wormholes in an emergent universe. }
\author{ Rikpratik Sengupta$^a$, Shounak Ghosh$^b$, B C Paul$^c$, M Kalam$^a\footnote{$^*$Corresponding author.\\
{\it E-mail addresses:} rikpratik.sengupta@gmail.com (RS), shounak.rs2015@physics.iiests.ac.in (SG), bcpaul@associates.iucaa.in (BCP), kalam@associates.iucaa.in (MK) }$.}

\address{$^a$Department of Physics, Aliah University, Kolkata 700 160, West Bengal, India.\\
         $^b$Department of Physics, Indian Institute of Engineering Science and Technology, Shibpur, Howrah 711103, West Bengal, India.\\
         $^c$Department of Physics, University of North Bengal, Siliguri 734 013, West Bengal, India..}

\begin{abstract}
	A non-singular Emergent Universe (EU) scenario within the realm of standard Relativistic physics requires a generalization of the Equation of State (EoS) connecting the pressure and energy density. This generalized EoS is capable of describing a composition of exotic matter, dark energy and cosmological dust matter. Since the EU scenario is known to violate the Null Energy Condition, we investigate the possibility of presence of static, spherically symmetric and traversable Lorentzian wormholes in an EU. The obtained shape function is found to satisfy the criteria for wormhole formation, besides the violation of the NEC at the wormhole throat and ensuring traversability such that tidal forces are within desirable limits. Also, the wormhole is found to be stable through linear stability analysis. Most ${importantly}$, the numerical value of the emergent universe parameter $B$ as estimated by our wormhole model is in agreement with and lies within the range of values as constrained by observational data in a cosmological context. Also, the negative sign of the second EU parameter $A$ as obtained from our wormhole model is in agreement with the one required for describing an EU, which further indicates on the existence of such wormholes in an emergent universe ${without}$ accounting for any additional exotic matter field or any modification to the gravitational sector.

\end{abstract}

\begin{keyword}
\texttt{Wormhole, Emergent Universe}
\end{keyword}

\end{frontmatter}

\section{Introduction}

It is well known today that the standard big bang cosmology is plagued by the singularity problem. The beginning of time, or the early stage of the universe cannot be described in the standard relativistic context as singularity appears in the Einstein field equations (EFE's). In fact, for describing the universe, as long as its radius does not exceed the Planck scale, a theory of Quantum Gravity (QG) is required. However, till date there is no single consistent QG theory that is fully developed, although plenty of work is going on in this direction and Loop Quantum Gravity (LQG)~\cite{Rovelli,Bojowald} along with the higher dimensional Superstring and M-theories~\cite{Polchinski,Gasperini} can be said to be the two main contenders in this aspect. The later also attempts to unify the four natural interactions. The problem with such QG theories is that they can be tested only in very extreme physical conditions, situations involving very high spacetime curvature, believed to be found inside the event horizon of black holes, or in the very early universe-at the moment of, or just following its creation.

Two decades back, Ellis and Maartens~\cite{EM} proposed a cosmological model known as the ``Emergent Universe", which attempted to resolve the long standing initial singularity problem within the classical context of Einstein's General Relativity (GR). As a positive curvature universe is not ruled out by observation, they argued that the role of a positive curvature term may be significant in the early universe, though it can be neglected at the late times. The consideration of a such a term results in a non-singular cosmology, with the universe originating as an Einstein's static universe (ESU) and is also free of the horizon problem. This static universe undergoes an inflationary phase and later reheating, to give the standard big bang era. Moreover, if the initial radius of the static universe is above the Planck scale, then a purely QG regime can be avoided altogether in this cosmological model.

This emergent universe scenario has attracted much attention from cosmologists in the last two decades~\cite{Bag,BCP1,AB1,AB2,UD,del,Beesh}. In a later work, Ellis~\cite{E2} extended the first model by considering the early universe to be dominated by a minimally coupled scalar field $\phi$, described by a physically interesting potential $V(\phi)=\bigg(Ae^{B\phi}-C\bigg)^2+D$, such that the constants $A$, $B$, $C$ and $D$ may be determined from specific properties of the emergent universe. In another work~\cite{E3}, it was shown that such a potential for the emergent universe can be reproduced by modifying the Lagrangian, adding a quadratic term of the scalar curvature, such that $L=R+\alpha R^2$, where the coupling parameter $\alpha$ turns out to be negative and the field can be identified with a negative logarithmic function of the curvature. A very important extension to the emergent universe (EU) scenario was done by Mukherjee et al.~\cite{M1}, where they obtained an EU even for a flat spacetime, in the context of the semi-classical Starobinsky~\cite{Starobinsky} model. A one parameter family of solutions describing the EU were obtained, such that parameter was determined by the number and species of the primordial fields.

In a following work, Mukherjee et al.~\cite{M2} obtained EU solution within the relativistic context, for a flat spacetime. They considered the composition of matter, such that the generalized Equation of State (EoS) is given as $p=A\rho-B\rho^\frac{1}{2}$. Such an EoS serves a three fold purpose. It accommodates the scope for description of the matter or the source of gravity sector by quantum field theory, besides the scope for including exotic matter capable of violating the energy conditions of GR in a cosmological context and also accommodating the present late time acceleration of the universe as inferred from supernova observations. For such an EoS, the universe is large enough initially to avoid a QG regime. For obtaining an emergent universe, we must have $B>0$. It may also be noted that an emergent universe is permitted with $A\leq0$~\cite{M2,B1}. For realistic values of the parameter $B$ constrained from observations~\cite{B2}, the universe is found to contain dark energy, exotic matter and cosmological dust (matter).

The idea of wormholes on a serious mathematical context date back to 1935, with the Einstein-Rosen bridge~\cite{ER} being proposed as a solution to the field equations of GR, which connected two different spacetime points located within the same universe, or even possibly in two different universes. Two decades later Misner and Wheeler~\cite{MW} revisited the idea and coined the term ``wormhole". However, such a wormhole was not traversable and found their existence from quantum fluctuations taking place in the spacteime foam. They are basically microscopic. The idea of Schwarzschild and Kerr wormholes cannot be considered seriously for traversability, as such wormholes contain event horizons. It is known that beyond an event horizon, the tidal forces are extremely large and also there is the presence of curvature singularity. The first formal, modern take on wormholes and their traversability was provided by Morris and Thorne in 1988~\cite{MT1}. In this view, a wormhole is considered as a topological as well as geometrical object characterized by a compact spacetime region which has a trivial boundary with a non-trivial interior and connects two spacetime points in the same universe or in two different universes. It has both local and global structure~\cite{Visserbook}. Such wormholes are well described by a static and spherically symmetric spacetime metric. Any closed time-like curve is known to violate causality and such a possibility may arise from traversable wormholes~\cite{MT2}.

Morris and Thorne~\cite{MT1} laid down a prescription for constructing a traversable wormhole. There are certain essential features that the radial metric component of a wormhole must follow in order to ensure traversability. We shall discuss these criteria in the context of our wormhole model in the concluding section. From a physical point of view, the tidal acceleration must be small enough so that any traveller attempting to traverse the wormhole does not get ripped apart. This can be ensured from the absence of horizons in the wormhole. Also, they suggested that the wormhole throat must be kept open so that it does not `pinch off', in order to ensure traversability through it. This is possible by the violation of the Null Energy Condition (NEC) at the throat. There are two ways in which the NEC may be violated. Firstly, by introduction of exotic matter at the throat within the relativistic context and secondly, by modifying the gravitational action, such that the additional terms in the modified field equations contribute effectively to the violation of NEC at the throat in the presence of ordinary matter. The possibility of a wormhole existing in the outer galactic halo region has been investigated recently~\cite{Kalam}. Wormholes have been studied in the modified gravity context~\cite{RS2,Dzhunushaliev,Bronnikov,Lobo2,Banerjee,Chakraborty2,Rahaman1} and also in the relativistic context with exotic matter~\cite{RS1,Barcelo,Hayward,Picon,Sushkov,Lobo,Zaslavskii,Chakraborty}. However, recently a successful traversable wormhole model has been proposed in the relativistic context~\cite{Konoplya} where the wormhole is supported by a Maxwell and two Dirac fields in the absence of any exotic coupling and exotic matter and the wormhole is asymmetric about the throat, such that there is no $Z_2$ or mirror symmetry.

As we have already discussed, in order to obtain an emergent universe solution with a flat spacetime, we have to either modify the Lagrangian describing the gravitational action or consider a generalized EoS for the matter source, within the relativistic context. Such an EoS is well known to accommodate exotic matter and  also accounts for the late-time acceleration of the universe. So, it could be natural that an emergent universe described by such an EoS accommodates traversable wormholes as well, due to such a composition of the universe. Lorentzian wormholes are the ones which exist on a Lorentzian spacetime manifold. Since experimental physics seems to favour a Lorentzian signature, here we will investigate the possibility of existence of static, traversable Lorentzian wormholes in the background of an emergent universe. In our previous papers, we had successfully constructed static and traversable Lorentzian wormholes involving higher dimensional braneworld gravity~\cite{RS2} or by introduction an additional tachyon field as the matter source~\cite{RS1}.  Time evolving Euclidean wormholes have been found to be present in an EU in the massive gravity context~\cite{PaulEU}. So, it would be worth an attempt to investigate the possibility of static and traversable Lorentzian wormholes being present in an EU without considering any additional matter field or modifying the Einstein-Hilbert action leading to standard relativistic GR.

In the following sections we have obtained the solutions for different physical parameters of the wormhole under the framework of the EU and investigated them different features of the wormhole that may be analyzed from the solution and consider its stability. A detailed analysis should throw light on whether a stable, traversable wormhole can exist automatically in an emergent universe setup.

\section{Mathematical model of the wormhole}

In this section we shall make use of the Einstein field equations (EFE's) to obtain the shape function for the wormhole and check whether it satisfies the criteria of being a well-defined function to describe the wormhole. We shall also check the validity of the null energy condition. The traversability condition can be verified by computing the tidal acceleration. The essential model parameters can by estimated from the Darmois-Israel junction conditions, from which we shall also compute the surface density and surface pressure of the wormhole. We shall also perform a linearized stability check in order to ensure that the wormhole is stable.

\subsection{The field equations}

For a static, spherically symmetric matter distribution, the spacetime is characterized by the line element

\begin{equation}
	ds^2=-e^{\nu(r)}dt^2+e^{\lambda(r)}dr^2+r^2(d\theta^2+sin^2\theta d\phi^2).
\end{equation}

For such a wormhole, the metric potential $e^\lambda(r)=\frac{1}{1-\frac{b(r)}{r}}$, where $b(r)$ represents the shape function of the wormhole. The Einstein Field Equations (EFEs) for such a wormhole in the relativistic context may be written down as (taking $G=$$c=$1)

\begin{eqnarray}
	&&\frac{b^\prime}{r^2} =8 \pi \rho,\label{eq6}\\
	&& \left(1-\frac{b}{r}\right)\left(\frac{\nu^\prime}{r}+\frac{1}{r^2}\right) -\frac{1}{r^2} =8 \pi p,\label{eq7}\\
	&&\left(1-\frac{b}{r}\right)\left(\nu'' + {{\nu^\prime}^2} +\frac{\nu^\prime}{r}  \right) -\frac{b^\prime -b}{2r}\left({\nu^\prime} +\frac{1}{r} \right) = 8 \pi p,\label{eq8}
\end{eqnarray}

As we are performing our analysis in the background of an Emergent Universe, we shall consider the static Lorentzian wormhole to be composed of dark energy, exotic matter and dust, represented by the generalized EoS~\cite{M2}.

\begin{equation}
	p \left( r \right) =A\rho \left( r \right) -B\sqrt {\rho \left( r \right) }.
\end{equation}

Here the parameters $A$ and $B$ characterize the Emergent Universe. For simplicity of our analysis we take the redshift function $\frac{\nu}{2}$ to be a constant which implies $\nu'=0$. This simplification can be found in case of other wormhole constructions with exotic matter in the relativistic context~\cite{RS1,Kartach}.

\subsection{Solution for the shape function}

Using the EoS in the matter conservation equation with a constant redshift function, the energy density of the wormhole is obtained as

\begin{equation}
	\rho \left( r \right) =\frac {B^2}{4 A^2}.
\end{equation}
We note that the energy density is expressed in terms of the EU parameters $A$ and $B$ as expected.

The shape function can be obtained by plugging in the energy density into Eq. (2) and turns out to be

\begin{equation}
	b(r) =\frac{2 \pi B^2 r^3}{3 A^2}+C_1.
\end{equation}

\begin{figure*}[thbp]
	\centering
	\includegraphics[width=7cm]{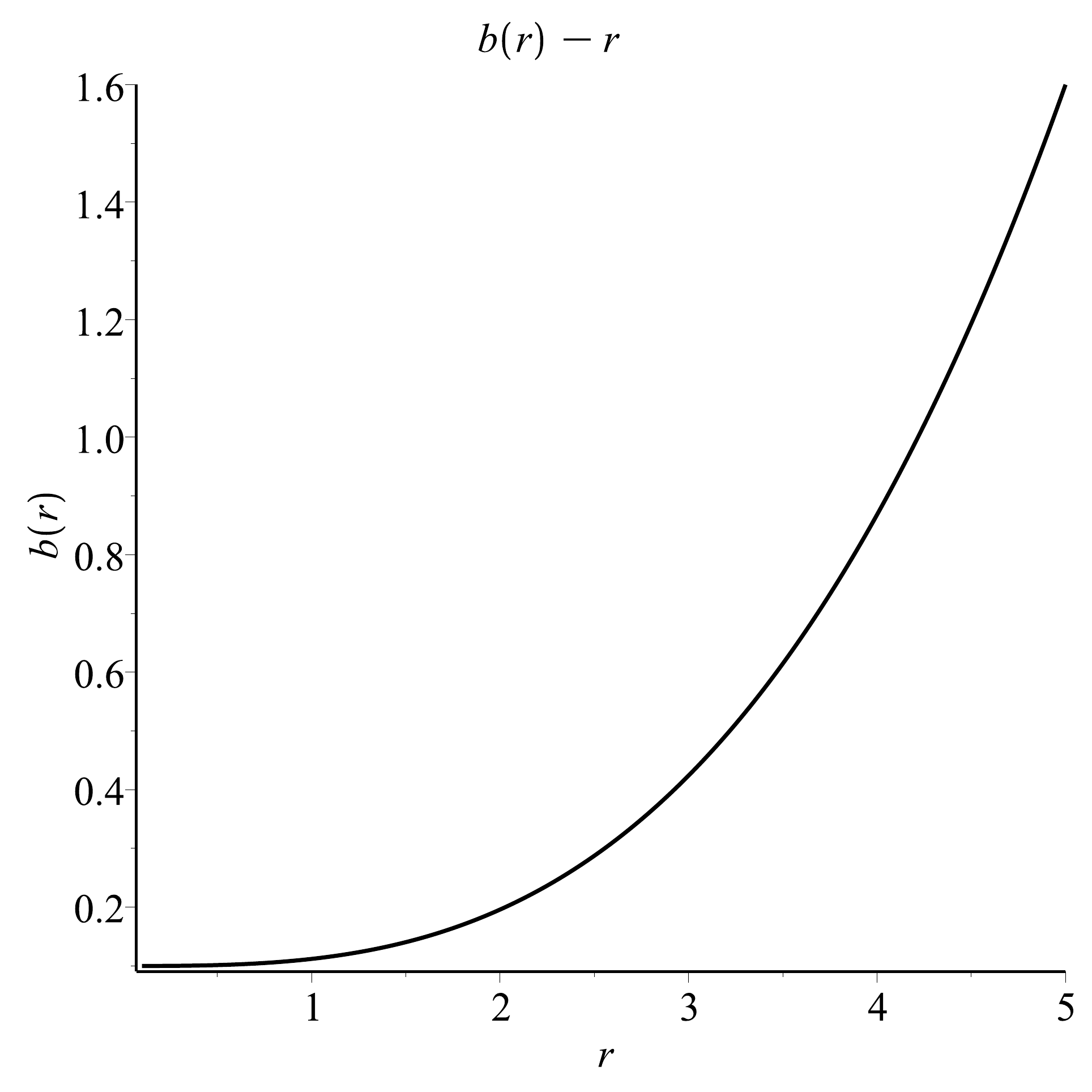}
	\caption{Variation of shape function with respect to $r$.}
\end{figure*}

We plot the shape function against $r$ in Figure 1. We have assumed that the throat radius of the wormhole is at $r=r_0=0.1$. As we can see from the Figure 1, the shape function satisfies the necessary criteria to describe a traversable wormhole as per the Morris Thorne prescription~\cite{MT1}. Firstly at the throat radius $r=r_0$, as we can see the shape function $b(r_0)=r_0$. Secondly, as $r$ becomes greater than $r_0$, at every point the shape function $b(r)<r$ for that point. So, the obtained shape function is satisfactory for describing a wormhole.

\subsection{Null Energy condition}

The null energy condition (NEC) in GR can be written in a simplified form as $\rho+p \geq 0 $. According to the Morris-Thorne prescription for construction of a traversable wormhole~\cite{MT1}, one of the most essential conditions for ensuring the traversability must be the violation of the NEC at the throat, so that the throat does not pinch off due to the gravitational attraction. In order to ensure this within the relativistic context, the matter constructing the wormhole must violate the NEC. So, for our model we must check whether the composite matter described by the generalized EoS in Eq. (6) can is capable of violating the NEC.

By solving the EFE in Eq. (4), we obtain the pressure to be given as

\begin{equation}
	p (r) =-\frac {2\,{B}^{2}\pi  r^3+3 C_1 A^2}{24 \pi A^2 r^3}.
\end{equation}
As expected, the pressure has a dependence on the EU parameters $A$ and $B$.

Summing up the above obtained pressure and energy density, we have

\begin{equation}
	p+\rho={\frac {4  B^2\pi   r^3-3 A^2 C_1}{24 \pi  A^2 r^3}}.
\end{equation}
\begin{figure*}[thbp]
	\centering
	\includegraphics[width=7cm]{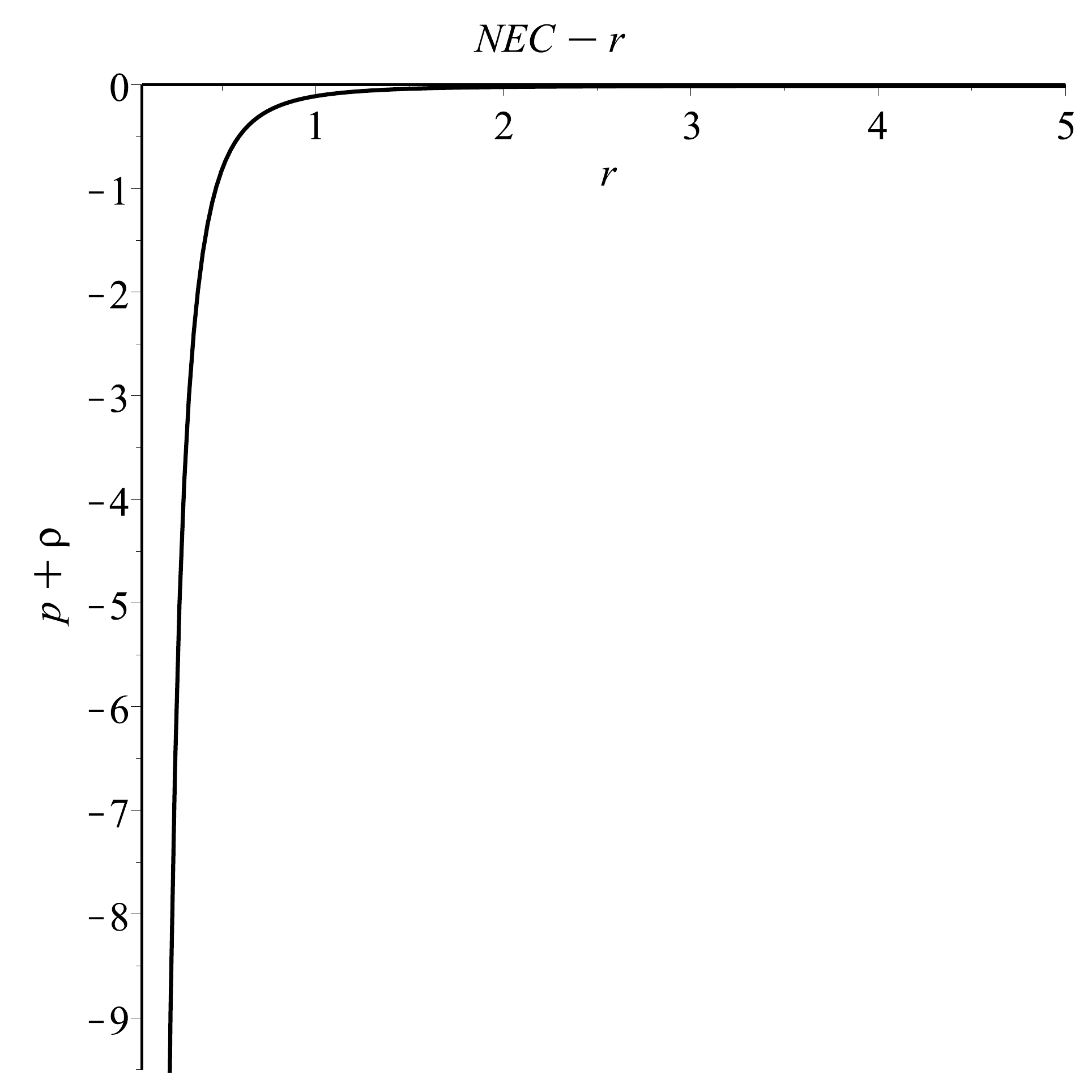}
	\caption{Variation of the NEC with respect to $r$.}
\end{figure*}

We plot $p+\rho$ as a function of $r$ in Figure 2. We see that the NEC is violated at the throat as $p+\rho$ turns out to be negative at the throat and as we further increase $r$, it continues to stay negative but with a vanishingly small value. It appears from the figure that $p+\rho$ turns zero after a value of $r$ close to 1.5, but on enlarging we can see that it continues to be negative with a vanishingly small value. This ensures that the flare out condition is satisfied which implies that the first derivative of the shape function  with respect to r, at the throat is less than unity. Hence, traversability is ensured.

\subsection{Darmois Israel Junction Condition}

\textbf{Outside the wormhole we consider a vacuum spacetime. So, the exterior to which the wormhole spacetime is to be matched can either be described by a Schwarzschild or deSitter metric depending on whether the vacuum has vanishing energy density and pressure ($\rho_{vac}=p_{vac}=0$) or constant energy density and pressure given by $p_{vac}=-\rho_{vac}=-\frac{\Lambda}{8\pi}$, respectively. However, for our analysis there is no need for considering a $\Lambda$-term (cosmological constant) as the composite  matter described by the EoS can describe the late time acceleration effectively even in the case of $\Lambda=0$. So, we consider a Schwarzschild exterior.  Thus, the spacetime exterior to the surface of the wormhole is described by the well known Schwarzschild metric given as}

\begin{equation}
	ds^2=-\left(1-\frac{2M}{r}\right)dt^2+\left(1-\frac{2M}{r}\right)^{-1}dr^2+r^2(d\theta^2+\sin^2 \theta d\phi^2).
\end{equation}

As a consequence of matter being present on the wormhole surface, there is an extrinsic discontinuity which produces an intrinsic surface energy density and surface pressure. The wormhole surface acts as the boundary between the interior and the exterior, as a result of which the wormhole structure is a geodesically complete manifold characterized by the EoS used in our model, which describes a composition of dark energy, exotic matter and cosmological dust. Being a geodesically complete manifold, the continuity of the metric coefficient at the wormhole surface yields a matching condition. We shall now obtain the components for the intrinsic stress-energy $S_{ij}$ using the Darmois-Israel prescription~\cite{Darmois,Israel}.

From the Lanczos equation~~\cite{Lanczos1924,Sen1924,Perry1992,Musgrave1996} we have

\begin{equation}
S_{j}^{i}=-\frac{1}{8\pi} (\kappa_{j}^{i}-\delta_{j}^{i} \kappa_{k}^{k}).
\end{equation}

The discontinuity in extrinsic curvature as a result of presence of matter on the wormhole surface is expressed as
\begin{equation}
	\kappa_{ij}=\kappa_{ij}^{+}-\kappa_{ij}^{-},
\end{equation}
and the extrinsic curvature is defined as
\begin{equation}\label{eq26}
	\kappa_{ij}^{\pm}=-n_{\nu}^{\pm}\left[\frac{\partial^{2}X_{\nu}}{\partial \xi^{i}\partial\xi^{j}}+
	\Gamma_{\alpha\beta}^{\nu}\frac{\partial X^{\alpha}}{\partial \xi^{i}}\frac{\partial X^{\beta}}{\partial
		\xi^{j}} \right]|_S,
\end{equation}
where $n_{\nu}^{\pm}$ is the unit normal vector which may be expressed in the form
\begin{equation}\label{eq27}
	n_{\nu}^{\pm}=\pm\left|g^{\alpha\beta}\frac{\partial f}{\partial X^{\alpha}}\frac{\partial f}{\partial X^{\beta}}
	\right|^{-\frac{1}{2}}\frac{\partial f}{\partial X^{\nu}},
\end{equation}
such that $n^{\nu}n_{\nu}=1$. $\xi^{i}$ denotes intrinsic coordinate of the wormhole surface and is described by the parametric equation $f(x^{\alpha}(\xi^{i}))=0$. Here $+$ and $-$ denotes the wormhole spacetime at the exterior and interior, respectively.

The surface stress energy tensor for the spherically symmetric line element has the components expressed as $S_{i}^{j}= diag(-\Sigma,\mathcal{P})$.
Here $\Sigma$ and $\mathcal{P}$ denotes the surface energy density and surface pressure respectively, at the wormhole surface $r=R$ which may be obtained as

\begin{eqnarray}
	\Sigma  &=&-\frac{1}{4\pi R}\bigg[\sqrt{e^\lambda}\bigg]_-^+\nonumber\\
	&=&\frac{1}{4\pi R}\left[\sqrt{\left(1-\frac{2M}{R}\right)} - \sqrt{1-\frac{b(r)}{r}} \right]=
    \frac{1}{4\pi R}\left[\sqrt{\left(1-\frac{2M}{R}\right)} - \sqrt{1-\frac{2 \pi B^2 r^2}{3A^2}-\frac{C_1}{r}}, \right]\\
    \\
\mathcal{P}  &=&\frac{1}{16\pi R } \bigg[\bigg(\frac{2f+f^\prime R}{\sqrt{f}}\bigg) \bigg]_-^+ \nonumber\\
&=&{\frac {1}{4\pi	R} \left( \sqrt {1-{\frac{2M}{R}}}-\sqrt{{1-\frac{C_1}{R}-\frac{2 B^2\pi R^2}{3A^2}}} \right)}+{\frac {M}{4 R^3\pi \sqrt{{1-\frac{2 M}{R}}}}}+\frac{\left(\frac{ B^2}{6 A^2}-\frac{C_1}{8 R^3 \pi}\right)}{ {\sqrt {{1-\frac{C_1}{R}-\frac{2 B^2\pi R^2}{3A^2}}}}}
\end{eqnarray}

As we are considering the construction of a static wormhole which does not evolve with time itself and is a local object in the EU, so at the surface of the wormhole the surface energy density and surface pressure must vanish ($\Sigma=\mathcal{P}=0$)\cite{RS2,Chakraborty}. \textbf{$\Sigma=0$ leads to the condition}

\begin{equation}
	b(r)|_{r=R}=2M,
\end{equation}
which yields
\begin{equation}
	- \left( -{\frac {2\pi  B^2{R}^{2}}{3A^2}}-{\frac {{C_1}}{R}} \right) R=2M.
\end{equation}

In addition, there is an additional boundary condition which involves the continuity of the metric potential $g_{rr}$
and its derivative $\frac{\delta g_{rr}}{\delta r}$ at the wormhole surface $r=R$. \textbf{It is to be noted that the same boundary condition obtained in Eq. (18) an Eq. (19) can also be arrived at by making use of the matching condition involving the continuity of the metric potential $g_{rr}$ to ensure smooth matching between the wormhole and exterior Schwarzschild spacetimes. Another matching condition involving the coninuity of the derivative of the radial derivative of the metric potential at the wormhole surface $r=R$, again in order to ensure smooth matching of the two spacetimes yield the second boundary condition which can be written as}
	
	\begin{eqnarray}
		&&\frac{\partial g_{rr}}{\partial r}|_{int}=\frac{\partial g_{rr}}{\partial r}|_{ext} \nonumber\\
		&\Rightarrow& -\frac{4 \pi B^2 R}{3A^2}+\frac{C_1}{R^2}=\frac{2M}{R^2}
\end{eqnarray}

\textbf{As there are three unknown model parameters (the EU parameters $A$ and $B$ and the integration constant $C_1$), we shall requaire an additional boundary condition. The third and final boundary condition comes from the vanishing of the surface pressure at the wormhole boundary, which is a consequence of the wormhole being static as already discussed. So, we can put Eq. (17) as zero which yields the third boundary condition. Now for the three unknown model parameters $A$, $B$ and $C_1$ we have three equations in the form of the three boundary conditions and we have evaluated the parameters by solving these equations using MAPLE.}
	
These boundary conditions allow us to evaluate the unknown model parameters $A$, $B$ both of which are the EU parameters
in addition to the integration constant $C_1$. We had previously assumed the throat radius to be $0.1$ (in $km$). As the
wormhole contains exotic matter as one of its ingredients, so it is appropriate to minimize the mass of the wormhole in
order to minimize the mass of exotic matter used. However, it may be noted here that we are not introducing any exotic
matter by hand. The EoS which supports an EU within the relativistic context is itself generalized to describe exotic
matter as a constituent. We assume the mass of the wormhole to be $0.8M_\odot$, which appears to be a realistic choice.
The surface or the boundary of the wormhole is taken at $R=5$ (in $km$).

Matching the junction conditions at the wormhole surface and using the values of the physical model parameters mentioned above,
we evaluate the unknown model parameters to be $A= -0.054$, $B = 0.003784713921$ and $C_1 = 0.09998799979$. These obtained values
have been used for the plots, and we shall justify their physical significance in the concluding section.

\subsection{Tidal acceleration}

For a traversable wormhole model, we must ensure that the tangential and radial components of the tidal acceleration for an observer
passing through the throat of the wormhole must be lesser than the acceleration due to gravity on earth, so that the observer does
not get ripped apart while traversing the throat due to large tidal gravitational forces.

The condition that has to be satisfied by the radial component of the tidal acceleration $|R_{rtrt}|$ in order to ensure traversability at the throat is

\begin{equation}
	|R_{rtrt}|=|(1-\frac{b}{r})\big[\frac{\nu''}{2}+\frac{\nu'^2}{4}-\frac{b'r-b}{2r(r-b)}.\frac{\nu'}{2}\big]|\leq g_{earth}.
\end{equation}

For our model, it has been considered that $\nu'(r)=0$. A constant redshift function is a convenient choice in case of many wormhole models with exotic matter in the context of standard GR~\cite{RS1,Kartach}. As a result, the radial component of the tidal acceleration vanishes and the condition of traversability is automatically satisfied for this component.

However, for the tangential component, the condition has to be evaluated at the throat as the tangential component is non-trivial and a constraint on the velocity of the traveller traversing the throat can be obtained on evaluating this condition. The condition for traversability is expressed by the inequality

\begin{equation}
	\gamma^2 |R_{\theta t \theta t}|+\gamma^2v^2 |R_{\theta r \theta r}|=|\frac{\gamma^2}{2r^2}\big[v^2(b'-\frac{b}{r})+(r-b)\nu'\big]|\leq g_{earth},
\end{equation}
where the left hand side represents the tangential tidal acceleration.

Plugging in the expression for $b(r)$ and using $\nu'(r)=0$ we get
\begin{equation}
\frac{{\gamma}^{2}}{2} \left| {\frac {v^2}{r} \left( -\frac{4}{3}{\frac {\pi B^2r}{A^2}}+{\frac {C_1}{r^2}} \right) } \right|\leq g_{earth}.
\end{equation}

Putting the values of the model parameters $B = 0.003784713920$, $A = -0.054$, $M = 0.8M_\odot$, $C_1 = 0.09998799990$ obtained from the matching conditions and taking $\gamma \approx 1$ (as the velocity of the traveller $v<<1$ so $\gamma=\frac{1}{\sqrt{1-v^2}} \approx 1$), the condition for traversability reduces to

\begin{equation}
\frac{1}{2} \left| {\frac {v^2}{r} \left( - 0.02400019199r+ 0.09998799990{r}^{-2} \right) } \right|\leq g_{earth}.
\end{equation}
At the throat of the wormhole, the abobe inequality reduces to
\begin{equation}
	v\leq 0.1414468192\sqrt{g_{earth}}.
\end{equation}

Since the upper limit on the velocity of the traveller traversing the throat of the wormhole turns out to be a realistic one for our wormhole, we claim that the tidal forces at the throat are not too strong to disrupt the traveller and hence traversability is ensured.

\subsection{Linearized Stability Analysis}

In order to perform a qualitative stability analysis of our obtained wormhole, it is assumed that the throat radius of the wormhole is a function of proper time. The throat radius is represented by $r_0=x(\tau)$. The energy density has the form
\begin{equation}
	\sigma=-\frac{1}{2\pi x}\sqrt{f(x)+\dot{x}^2},
\end{equation}
and the pressure may be expressed as
\begin{equation}
	p=\frac{1}{8\pi}\frac{f'(x)}{\sqrt{f(x)}}-\frac{\sigma}{2},
\end{equation}

such that $f(x)=1-\frac{2M}{x}$. Here $M$ denotes the wormhole mass.

Making use of the conservation equation, an equation of motion
\begin{equation}
	\dot{x}^2+V(x)=0,
\end{equation}
can be obtained, such that the potential $V(x)$ can be expressed as
\begin{equation}
	V(x)=f(x)-[2\pi x \sigma (x)]^2.
\end{equation}

The objective now is to perform stability analysis considering linearization around $x_0$, which is an assumed static solution to Eq.(27).

Taylor expanding the potential around the assumed static solution $x_0$ yields \begin{equation}
	V(x)=V(x_0)-V'(x_0)(x-x_0)+\frac{1}{2}V''(x_0)(x-x_0)^2+O[(x-x_0)^3],
\end{equation}
'prime' indicating derivative with respect to x.

Since the wormhole spacetime considered by us is static, we have $V(x_0) = 0$ and
$V'(x_0) = 0$.  Thus to ensure stability of the wormhole, it is necessary that  $V''(x_0) > 0$. A parameter $\beta$ is introduced, physically representing the sound speed which is defined as
\begin{equation}
	\beta=\frac{\delta p}{\delta \sigma}.
\end{equation}

The second derivative of the potential can be expressed in terms of the newly defined physical parameter $\beta$ as
\begin{equation}
	V''(x)=f''(x)-8 \pi^2 [ (\sigma +2p)^2+ \sigma (\sigma+p)(1+2\beta).
\end{equation}

Expressing the stability condition for the wormhole  $V"(x_0) > 0$ in terms of $\beta$, we have
\begin{equation}
	\beta< \frac{\frac{f''(x_0)}{8\pi^2}-(\sigma +2p)^2-2\sigma(\sigma+p)}{4 \sigma (\sigma +p)}.
\end{equation}

Plugging in the expressions for $\sigma$ and $p$, the stability criterion for the wormhole in terms of $\beta$ takes the final form

\begin{equation}
	\beta< \frac{x_0^2 (f_0')^2-2x_0^2 f_0'' f_0}{4 f_0(a_0 f_0' -2f_0)}-\frac{1}{2}.
\end{equation}

The parameter $\beta$ can be calculated for our wormhole model as
\begin{eqnarray}
	\beta={\frac {-2{x_0}^{6}+ \left( -12\pi +9 \right) m{x_0}^{5}+20{m}^
			{2} \left( \pi -\frac{1}{2} \right) {x_0}^{4}}{8{x}^{5} \left( -x_0+2 m \right)
			\pi  \left( -x_0+3 m \right) }}.
\end{eqnarray}

We plot $\beta$ for different values of $x_0$ in Figure 3. The regions denoted as 1, 2 and 3 satisfy the stability criterion. Thus, we can claim that the wormhole model obtained by us in the background of an EU is stable.

\begin{figure*}[thbp]
	\centering
	\includegraphics[width=7cm]{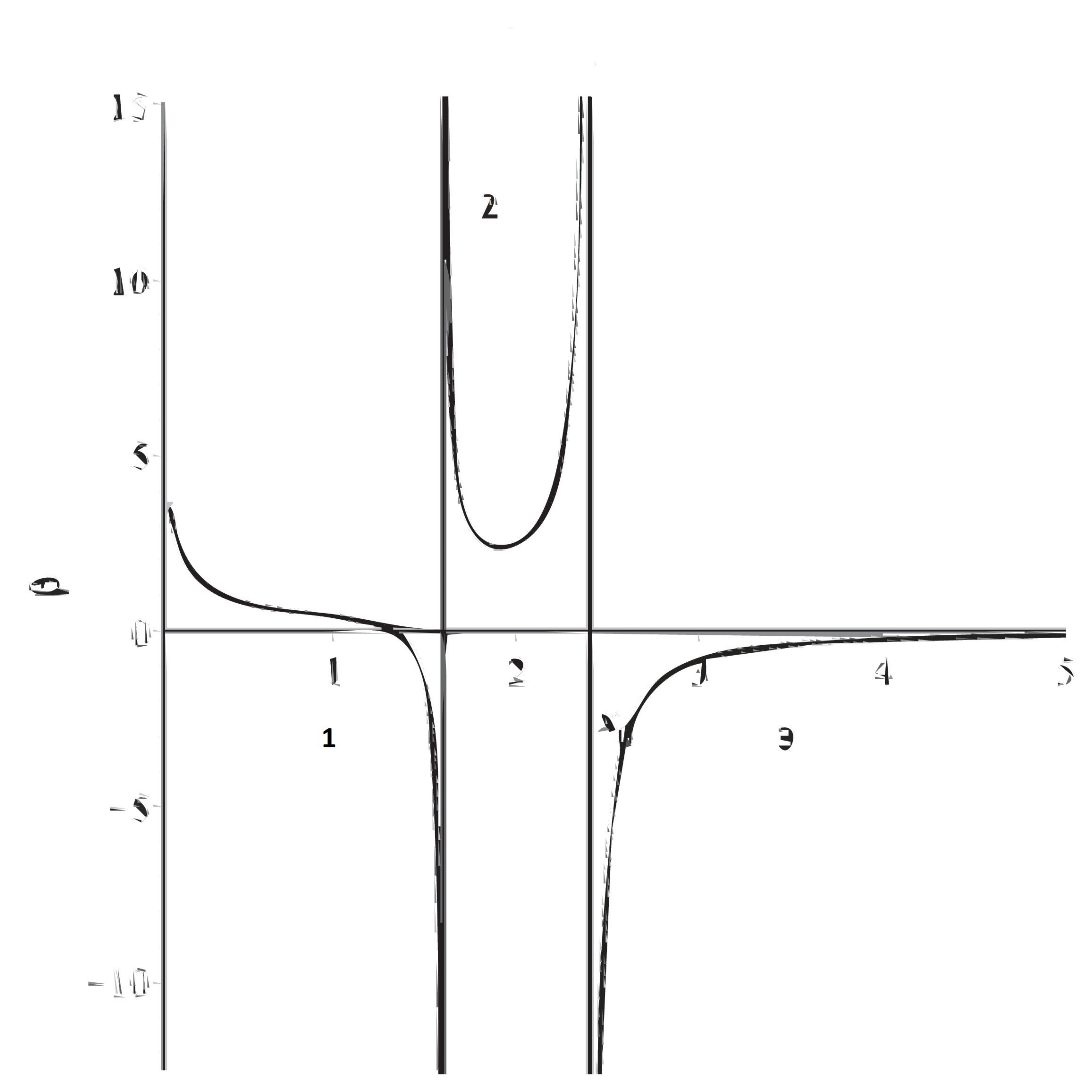}
	\caption{Variation of $\beta$ with respect to $x_0$.}
\end{figure*}

\section{Discussion and Conclusion}

In this paper, we explore the possibility of constructing static, traversable and spherically symmetric Lorentzian wormholes in the background of an Emergent Universe (EU). As we have discussed in the introductory section, EU cosmological models have been widely explored by cosmologists due to the absence of the initial singularity within the relativistic context. It is the most well established non-singular cosmological picture which is consistent even without modifying General Relativity. However, in order to accommodate EU in the relativistic context, a generalized EoS is introduced which describes an universe composed of exotic matter, dark energy and cosmological dust for suitable values of the model parameters. Again, it is known that in order to have an EU, it is necessary that the NEC must be violated~\cite{Zhu}. Also, this is a necessary condition for the successful construction of a traversable wormhole~\cite{MT1}. In the relativistic context, exotic matter also becomes effective in constructing traversable wormholes. Therefore, the common grounds of requirement for exotic matter and violation of the NEC make it worth an attempt to explore the possibility of existence of traversable wormholes in an Emergent Universe. Also, the existence of time evolving  Euclidean wormholes have been previously found in EU in the context of massive gravity~\cite{PaulEU}. This provides an additional motivation for the present investigation.

We have considered a constant redshift function to simplify our analysis. Using the generalized EoS used to describe an EU, we obtain a solution for the shape function and on plotting the shape function, it turns out that all the essential criteria that are required for a shape function to describe a traversable wormhole have been obeyed. At the throat radius $r_0$, the shape function takes the value of the throat radius itself. Moreover, for all values of $r>r_0$, the ratio remains $\frac{b(r)}{r}<1$ (from Fig. 1).

The energy density and pressure inside the wormhole are obtained making use of the field equations and the conservation equation. They are found to depend on the EU parameters. Plotting the variation of $p+\rho$ along $r$, we find that the NEC is violated at the throat of the wormhole. This ensures that the derivative of the shape function $b(r)$ with respect to $r$ must be less than unity at the throat, which is known as the flaring-out condition. Satisfying this condition physically means that the throat of the wormhole does not get pinched off due to the gravitational attraction, thus preventing the wormhole structure from collapsing at the throat. This can be justified by the presence of exotic matter and dark energy components in the EU, described by the generalized EoS.

The discontinuity in the extrinsic curvature resulting from the presence of matter at the surface of the wormhole, in turn results in the origin of a surface stress-energy term containing surface energy density and surface pressure terms as its components. These components have been obtained and are also found to be dependent on the EU parameters. Since we have performed our analysis for a static wormhole metric, so we consider the surface density and pressures to vanish at the junction. This results in yielding the first boundary condition. Other boundary conditions can be obtained from the continuity of the metric potential and its derivative at the junction.

These matching conditions at the surface of the wormhole enable us to obtain an estimate of the unknown model parameters, namely the EU parameters $A$ and $B$ and also the unknown integration constant $C_1$, by choosing realistic values of the parameters associated with the wormhole. The throat radius and the boundary of the wormhole are considered to be at $0.1 km$ and $5 km$, respectively. These are not unique choices but the values we have considered are reasonable. Since, exotic matter is one of the components of an EU, the wormhole in consideration will also contain some exotic matter as one of its constituents. So, it is justified to minimize the mass of the wormhole although dust matter is also a constituent. We make a reasonable assumption that the mass of the wormhole is $0.8M_\odot$. Using the chosen values of the above mentioned wormhole parameters, we get the integration constant $C_1=0.09998799979$. This is of relatively less physical significance.

However, it is interesting to check the values we obtain for the EU parameters, specially the parameter $B$ as this parameter has been constrained observationally in a cosmological context~\cite{B2}. If the value we obtain for the parameter  $B$ from our wormhole model happens to lie within or in close approximate to the constrained observational range, it would provide strong support in favor of existence of static traversable Lorentzian wormholes in an EU. Also, for EU, the parameter $A$ must be negative and $B$ must be slightly positive~\cite{M2}. However when we applied the EoS containing these two parameters to the field equations to obtain the wormhole solution, we did not put any constraint on the sign of these parameters by hand. So the obtained negativity and positivity of $A$ and $B$, respectively is an indication in favor of the possibility of existence of such wormholes in an EU. The observationally constrained value for the parameter $B$ for an EU constituted of exotic matter, dark energy and matter is $0.003 < B < 0.5996$~\cite{B2}. From our model, the obtained value from the matching condition is $B = 0.003784713921$, which is within the constrained range. This, along with the small negative value obtained for the parameter $A$, essential to construct an EU, provides strong support in favor of the existence of wormholes in an EU.

It is also essential to check that the tidal force at the wormhole throat is not strong enough, such that an observer attempting to traverse the throat of the wormhole gets ripped apart due to the tidal forces. The radial component of the tidal force becomes insignificant with the vanishing of the first derivative of the redshift function. However, the tangential component has a contributing term even for constant $\nu$. Considering the upper limit for the tidal acceleration to be equal to the acceleration due to gravity on the earth, we obtain a reasonable upper limit on the velocity with which the traveller can attempt to traverse the wormhole throat safely, without getting ripped apart due to excess tidal forces. Finally, we perform a linearized stability analysis to check whether our obtained wormhole solution in an EU is stable. Plotting the parameter $\beta$ with respect to $x_0$, we show three regions of stability which justifies that the wormhole we have obtained is a stable one. The stable regions are obtained imposing constraint on the sound speed in terms of the wormhole solution parameters.

There have been some recent investigations on the possibilities of detecting wormholes~\cite{DS,Torres,KonoZhi,Churilova}. As we know, wormholes are asymptotically flat tubular structures that connect two different spacetimes where the individual fluxes might not be conserved, as a result of which there shall be mutual observable effects on objects present within close distances of both the wormhole mouths~\cite{DS}. It may be possible to detect such effects on the orbits of stars in close approximately to the black hole present at the centre of our galaxy. Wormholes may also produce some micro-lensing effects which might appear identical to gamma ray bursts and these effects can lead to obtaining a higher limit on the mass density of wormholes making use of the BATSE data~\cite{Torres}. The scattering properties near a traversable and rotating wormhole can be explored considering the quasinormal ringing of black holes. Asymmetric wormholes shall be characterized by super radiance, while the wormholes exhibiting symmetry differ from black holes due to non-identical simultaneous ringing at a number of dominant multipoles~\cite{KonoZhi}. If the redshift function of the wormhole is variable, the radial tidal forces do not vanish, resulting in possible detection of long-lived quasinormal modes dubbed as 'quasi-resonances' in the background of the wormhole~\cite{Churilova}.

We may conclude that not only time evolving Euclidean wormholes can exist in the EU scenario in the massive gravity context, but also static, traversable Lorentzian wormholes can also be accommodated in an EU within a relativistic context. The EoS describing an EU in the relativistic context is capable of producing stable, static, traversable wormhole solutions violating the NEC. Moreover, the sign of the EU parameter $A$ is obtained as desired for an EU and also the numerical value of the EU parameter $B$ appearing for a consistent wormhole model, is estimated to be within the range of values constrained from an observational basis in a cosmological context. Therefore, it can be concluded that Lorentzian wormholes are naturally present in an Emergent Universe without taking into account either any additional matter field (like our previous wormhole model supported by a tachyonic field~\cite{RS1}) or any modification to standard GR (like our previous wormhole model in the braneworld gravity context~\cite{RS2}).

\section{Acknowlwdgement}

MK and BCP is thankful to the Inter-University Centre for Astronomy and Astrophysics (IUCAA),
Pune, India for providing the Visiting Associateship under which a part of this work was carried out.
RS is thankful to the Govt. of West Bengal for financial support through SVMCM scheme.
SG is thankful to the Directorate of Legal Metrology under the Department of Consumer Affairs, West Bengal for their support.

\end{document}